\documentclass[aps,prx,twocolumn,longbibliography,footinbib,superscriptaddress]{revtex4-2}
\usepackage{graphicx}
\usepackage{amssymb, amsmath,mathtools}
\usepackage{bbm}
\usepackage{enumerate}
\usepackage{hyperref}
\usepackage{dsfont}
\usepackage{physics}
\usepackage[version=4]{mhchem}
\usepackage{xcolor}
\usepackage{dcolumn}
\usepackage{wasysym} 
\usepackage{xspace}
\usepackage[normalem]{ulem}
\usepackage{newtxtext,newtxmath}

\newcommand{\ve}[1]{\boldsymbol{#1}}
\usepackage{orcidlink}

\graphicspath{{./}{./Figures/}}

\newcommand{\subref}[2]{\ref{#1}\hyperref[#1]{#2}}

\hypersetup{colorlinks=true,linkcolor=blue,citecolor=blue,urlcolor=blue}

\begin{document}
	
	\title{Magnetic-Field-Driven Dimensional Reduction in a Quantum Antiferromagnet}
	\author{Subhankar Khatua\,\orcidlink{0000-0002-5439-7504}}
	\affiliation{Institute for Theoretical Solid State Physics, IFW Dresden and W\"urzburg-Dresden Cluster of Excellence ctd.qmat, 01069 Dresden, Germany}	
	\affiliation{\mbox{Institut f\"ur Theoretische Physik und Astrophysik,
			Universit\"at W\"urzburg, 97074 W\"urzburg, Germany}}
	\author{Marcin Raczkowski\,\orcidlink{0000-0003-1248-2695}}
	\affiliation{\mbox{Institut f\"ur Theoretische Physik und Astrophysik,
			Universit\"at W\"urzburg, 97074 W\"urzburg, Germany}}
			
	\author{Jeroen van den Brink\,\orcidlink{0000-0001-6594-9610}}
	\affiliation{Institute for Theoretical Solid State Physics, IFW Dresden and W\"urzburg-Dresden Cluster of Excellence ctd.qmat, 01069 Dresden, Germany}
	\affiliation{Institute for Theoretical Physics, Technische Universit\"at Dresden, 01062 Dresden, Germany}

	\author{Fakher F. Assaad\,\orcidlink{0000-0002-3302-9243}}
	\affiliation{\mbox{Institut f\"ur Theoretische Physik und Astrophysik,
			Universit\"at W\"urzburg, 97074 W\"urzburg, Germany}}
	\affiliation{W\"urzburg-Dresden Cluster of Excellence ctd.qmat, Am Hubland, 97074 W\"urzburg, Germany}

\date{\today}
	
\begin{abstract}
Low dimensionality enhances quantum fluctuations, triggering novel states of quantum matter to emerge.
In real materials, low dimensionality usually arises from spatially strongly anisotropic couplings. 
Here, we demonstrate a different mechanism: in two-dimensional systems with coupled alternating ferromagnetic (FM) and antiferromagnetic
(AFM) spin-$1/2$ chains, an applied magnetic field may drive a dimensional reduction. 
Under magnetic  field, the FM chains polarize and stiffen, suppressing 
the propagation of transverse AFM fluctuations from one chain to another, and effectively induce one-dimensional behavior at low energies. 
For a model describing botallackite, \ce{Cu2(OH)3Br}, quantum Monte Carlo dynamics show that beyond a critical magnetic field, 
the low-energy spectrum reduces to that of a one-dimensional AFM Heisenberg spin-$1/2$ chain with 
field-dependent incommensurate two-spinon fluctuations, providing clear signatures for inelastic neutron scattering. 
\end{abstract}

	\maketitle

A central goal in condensed matter physics is to realize novel states of quantum matter in real materials. 
For this, low-dimensional quantum magnets offer an attractive route, as their emergent quantum properties are fundamentally distinct from those of long-range ordered higher-dimensional systems. 
Prominent examples include fractionalized spinon excitations in one-dimensional Heisenberg antiferromagnetic spin-$\frac{1}{2}$ chain~\cite{Bethe1931,Giamarchi} and Majorana excitations in a Kitaev spin liquid on the honeycomb lattice~\cite{Kitaev2006}. 
However, ideal realizations of strictly one- or two-dimensional models are challenging, since real materials are inherently three-dimensional~\footnote{with the exception of magnetic van der Waals materials~\cite{vanderWaals2018}}. Nevertheless, effectively one-dimensional physics can emerge in anisotropic systems where the intrachain coupling is dominant,  and  at temperature scales that exceed the interchain coupling. In such cases, correlations develop along the chains while interchain correlations remain incoherent. 
Generically, there will be a temperature scale below which the system exhibits higher-dimensional behavior, so that the aforementioned dimensional reduction does not constitute a quantum phase transition. 
We note however that exceptions to this generic expectation have been discussed in experiments, in the framework of the renormalization group, and in model calculations \cite{Batista2007,Sebastian2006,Matthias2007,Heinze25,Martin25}. 
\begin{figure}[h!]
        \includegraphics[width =0.92\columnwidth]{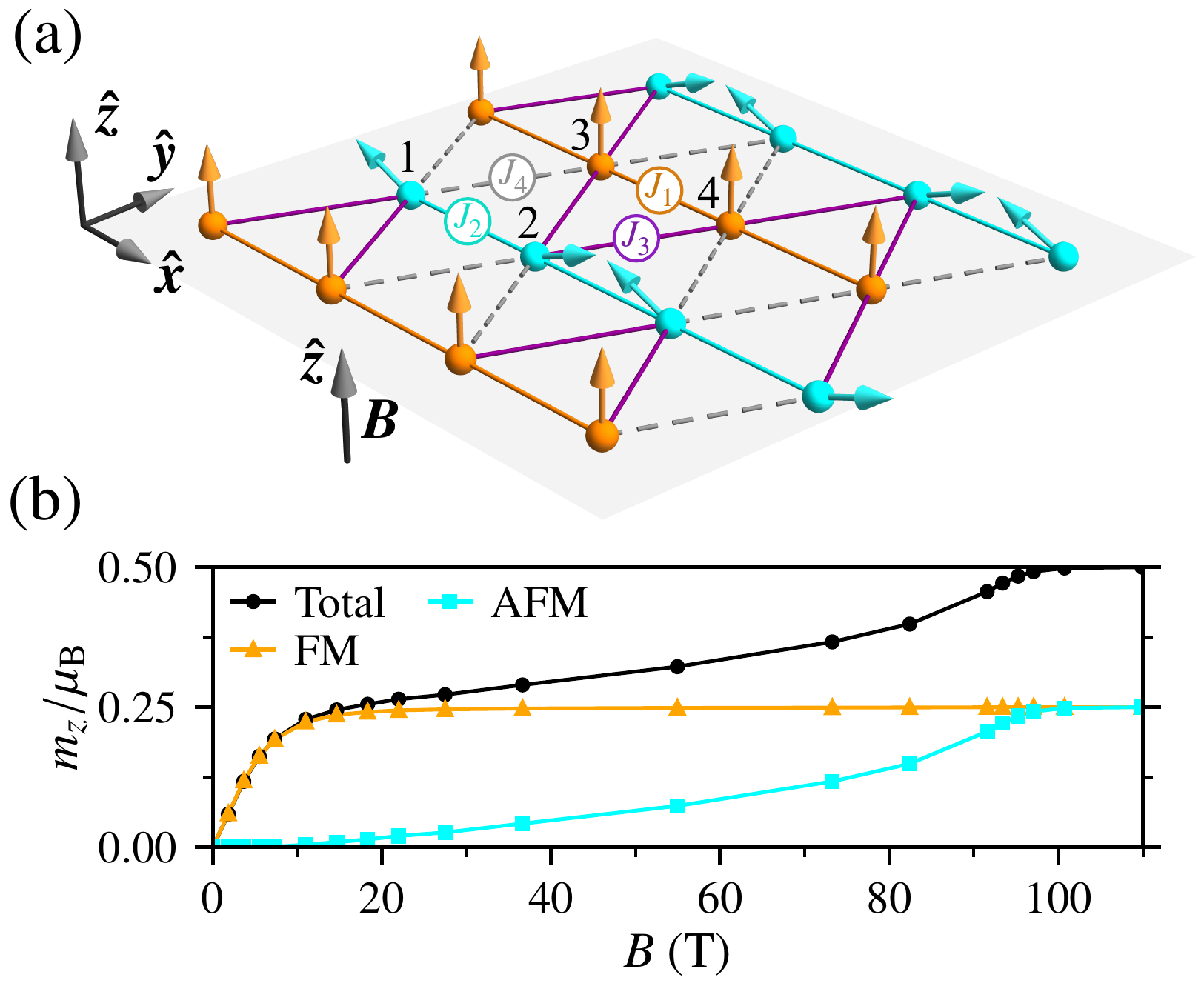}
        \caption[]{ (a) Schematic of the two-dimensional model for botallackite, consisting of alternating FM ($J_1$) and AFM ($J_2$) Heisenberg spin-$\frac{1}{2}$ chains in the presence of an out-of-plane magnetic field. The chains are coupled via interchain AFM Heisenberg interactions $J_3$ and $J_4$. Above the critical magnetic field, the FM chains are nearly fully polarized, while spins on the AFM chains remain canted. (b) Magnetization as a function of magnetic field obtained from QMC simulations at $T = 3.074~\mathrm{K}$. The total magnetization per spin, together with the separate FM and AFM chain contributions, is shown.} 
        \label{fig.model}
\end{figure}
Representative examples of dimensional reduction include quasi-one-dimensional Heisenberg spin-$\frac{1}{2}$ chain compounds such as \ce{KCuF3}\cite{Nagler1991,Tennant1993,Lake13}, \ce{BaCu2Si2O7}~\cite{Zheludev2000}, and \ce{CuF2(D2O)2(pyz)}\cite{Skoulatos2017}, 
weakly coupled spin ladder systems \ce{CaCu2O3}~\cite{Lake2010} and \ce{Cu3SO4(OH)4}~\cite{Kulbakov2022}.

 A natural question is whether dimensional reduction may also emerge in systems without spatially strongly anisotropic couplings. 
 More specifically, can an external tuning parameter drive a higher-dimensional quantum magnet into an emergent lower-dimensional regime? Furthermore, can such emergent dimensional reduction be realized in a material?
  
\begin{figure*}
        \centering
        \includegraphics[width = \textwidth]{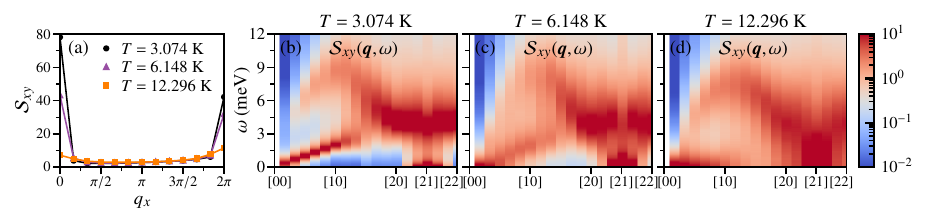}
        \caption[]{(a) Static structure factor for various temperatures at zero magnetic field for a path $(q_x,q_y) = (0,\pi) \rightarrow (2\pi,\pi)$.  (b)--(d): Dynamical structure factors for the same temperatures at zero magnetic field. $\ve{q}=[00]$ corresponds to the total spin. Since this is a conserved quantity, the dynamical structure factor exhibits a Dirac delta peak at $\omega =0$. We thus leave [00] blank.
        } 
        \label{fig.dyn-sfac-temp}
\end{figure*} 
Recently, magnetization and thermodynamic measurements on the minerals botallackite, \ce{Cu2(OH)3Br}~\cite{Reinold2025},  and atacamite, \ce{Cu2Cl(OH)3}~\cite{Heinze25}, have suggested the presence of a possible \emph{magnetic-field-driven} dimensional reduction. 
For botallackite, a \textit{quasi}-magnetization plateau has been observed experimentally under applied magnetic field, and quantum Monte Carlo (QMC) studies of the corresponding spin model have proposed an interpretation in terms of emergent quasi-one-dimensional behavior~\cite{Reinold2025}. 
The material consists of alternating ferromagnetic and antiferromagnetic spin-$\frac{1}{2}$ chains coupled by sizable interchain interactions~\cite{HZhang20,Gautreau2021,Povarov2024,Reinold2025}. 
Under an applied magnetic field, the ferromagnetic chains can rapidly polarize, whereas the antiferromagnetic chains are described by a canted antiferromagnetic state~\cite{Povarov2024,Reinold2025}.  Furthermore, Ref.~\cite{Reinold2025} suggests that since the transverse  spin fluctuations of the antiferromagnetic chains are orthogonal to the polarization of the ferromagnetic chain,  they cannot propagate  from one chain to another. 
While this study provides indications for dimensional reduction, it remains indirect, as it does not consider any underlying dynamics.

In this Letter, we provide direct theoretical evidence for this magnetic-field-driven dimensional reduction through large-scale QMC calculations of dynamical correlations in the corresponding two-dimensional Heisenberg spin model. 
We show that beyond a threshold magnetic field, the dynamical structure factor evolves into that of a one-dimensional spin-$\frac{1}{2}$ Heisenberg antiferromagnetic chain characterized by a gapless spinon continuum. 
At higher fields, the continuum undergoes a characteristic field-dependent reconstruction consistent with a partially polarized Heisenberg spin-$\frac{1}{2}$ chain. Our results establish magnetic-field control as a general route to emergent dimensional reduction in quantum magnets, and provide clear spectroscopic signatures accessible to inelastic neutron scattering experiments.

\textit{Model and Method}---Botallackite is a quasi-two-dimensional quantum magnet whose magnetic properties are governed by Cu$^{2+}$ ions with spin-$\frac{1}{2}$ moments arranged on a square lattice~\cite{Zhao2019,HZhang20,Gautreau2021,Povarov2024,Reinold2025}.  
As shown in Fig.~\subref{fig.model}{(a)}, the system consists of alternating ferromagnetic (FM) and antiferromagnetic (AFM) Heisenberg spin chains coupled by interchain interactions.  
The intrachain couplings on the FM and AFM chains are denoted by $J_1$ and $J_2$, respectively, while $J_3$ and $J_4$ represent the interchain couplings.
 We therefore consider the following Hamiltonian in the presence of an out-of-plane magnetic field $B\ve{\hat{z}}$,
\begin{equation}
H =\sum_{\ve{r},\ve{r}',\alpha,\beta} J_{\ve{r}\alpha,\ve{r}'\beta}\ve{S}_{\ve{r}\alpha}\cdot\ve{S}_{\ve{r}'\beta} - g\mu_{\rm B} B\sum_{\ve{r},\alpha}S^z_{\ve{r}\alpha},
\label{eq.model}
\end{equation}
 where $\ve{r}, \ve{r}'$ denote the positions of the unit cells on the square lattice, and $\alpha, \beta = 1,\cdots, 4$ label the four orbitals within each unit cell. The orbital positions are $\ve{r}_1 = (0,0)$, $\ve{r}_2 = (\frac{1}{2},0)$, $\ve{r}_3 = (\frac{1}{4},\frac{1}{2})$, $\ve{r}_4 = (\frac{3}{4},\frac{1}{2})$, and  
 $\ve{S}_{\ve{r}\alpha}\equiv \left(S_{\ve{r}\alpha}^x, S_{\ve{r}\alpha}^y,S_{\ve{r}\alpha}^z \right)$ represents a spin-$\frac{1}{2}$ operator at $(\ve{r}+\ve{r}_\alpha)$. 
 Orbitals  $1$ and $2$ belong to the AFM chains, while $3$ and $4$ belong to the FM chains.  
 The exchange coupling $ J_{\ve{r}\alpha,\ve{r}'\beta}$ takes values $J_1, J_2, J_3$, or $J_4$ depending on the bond, as shown in Fig.~\subref{fig.model}{(a)}, with $g$ the Land\'e-$g$ factor and $\mu_{\rm B}$ the Bohr magneton. 
 The experimentally estimated values are  $J_2 = 5.3$ meV, $J_1 = - 0.3\,J_2$, $J_3= 0.2\,J_2$, $J_4\approx 0$, and $g = 2.24$~\cite{HZhang20,Reinold2025}.   
The model considered here retains only SU(2)-symmetric spin interactions, and omits easy-axis anisotropy, dominant in the FM chains, as well as Dzyaloshinskii-Moriya interaction terms as both are negligible compared to the SU(2)-symmetric exchange terms~\cite{HZhang20}.

To solve the model we use the  Algorithms for Lattice Fermions (ALF)~\cite{ALF_v2.4}  implementation of the  finite-temperature auxiliary-field QMC simulations \cite{Blankenbecler81,White89,Assaad08_rev}, which is numerically exact and free of the negative-sign problem for the unfrustrated model of Eq.~\ref{eq.model}. 
In particular, we adopt an Abrikosov fermion representation of the spin-$\frac{1}{2}$ operators, and decouple the resulting four-fermion interactions using a Hubbard-Stratonovich transformation. The constraint of single fermion occupancy per site is enforced by a Hubbard interaction~\cite{SatoT21,HZhang20}. The  analytical continuation was
carried out  using the ALF \cite{ALF_v2.4} implementation of the  stochastic analytical continuation algorithm of Refs.~\cite{Sandvik98,Beach04a,Shao23}. 
Simulations have been carried out on a lattice of $12\times 6$ unit cells, each containing four orbitals  with Trotter step $\Delta \tau\,J_2= 0.1$. 
Further details of the QMC simulations are provided in the Supplemental Material~\cite{SM}.

First, we compute the net magnetization per spin using QMC for several magnetic field strengths at low temperature $T = 3.074~\mathrm{K}$, as shown in Fig.~\subref{fig.model}{(b)}. 
This reproduces the magnetization curve reported in Ref.~\cite{Reinold2025}. 
As can be seen in Fig.~\subref{fig.model}{(b)}, the net magnetization rises rapidly at low fields. Above a critical field $B_c \approx 16~\mathrm{T}$, it enters a regime of much slower growth, forming a plateau, and eventually reaches full polarization at $B_s \approx 97~\mathrm{T}$. 
The initial increase in magnetization originates from the quick polarization of the FM chains [see Fig.~\subref{fig.model}{(b)}]. 
In the plateau regime, the FM chains are already fully polarized, whereas the magnetization of the AFM chains continues to increase gradually with increasing field. 
Dimensional reduction to decoupled AFM chains has been proposed in this plateau regime. 
Finally, once the AFM chains become fully polarized, the entire system reaches saturation with all spins aligned along the $\ve{\hat{z}}$ direction.

Next, we compute the spin susceptibility of the FM and AFM chains in the longitudinal ($\mu = z$) and transverse ($\mu = x, y$) directions, defined as 
\begin{equation}
	\chi^{c}_{\mu}(\ve{q},\omega) =    i \int_{0}^{\infty} dt e^{i\omega t} \langle [\hat{S}^{c}_{\mu}(\ve{q},t), \hat{S}^{c}_{\mu}(-\ve{q},0)] \rangle,
	\label{eq.susceptibility}
\end{equation}
where $c=$ AFM or FM, corresponding to the AFM or FM chains and $\langle[\cdots] \rangle$ denotes the thermal average of the commutator.  
For the AFM chains, we define $\hat{S}^{\rm AFM}_{\mu}(\ve{q}) = \frac{1}{\sqrt{N}}\sum_{\ve{r}, \alpha= 1,2 }
 e^{i\ve{q}\cdot (\ve{r} + \ve{r}_\alpha)} S^{\mu}_{\ve{r}\alpha}$, where $N$ is the total number of unit cells in the system.  
 A similar definition applies for the FM chains with $\alpha = 3,4$. 
 The  dynamical structure factor reads as  
 \begin{equation}
	\mathcal{S}^{c}_{\mu}(\ve{q},\omega) = \frac{1}{\pi}\frac{\mathrm{Im} \chi^{c}_{\mu}(\ve{q},\omega)}{1-e^{-\beta \omega}}, 
 \end{equation}
 where $\beta^{-1} = k_{\rm{B}}T$, and the static structure factor can be obtained as $\mathcal{S}^{c}_{\mu}(\ve{q}) = \int d\omega \mathcal{S}^{c}_{\mu}(\ve{q},\omega)$. 
 Due to the in-plane U(1) symmetry of the Hamiltonian, we will write the transverse dynamical structure factor as $\mathcal{S}^{c}_{xy}(\ve{q},\omega) \equiv  [\mathcal{S}^{c}_{x}(\ve{q},\omega) + \mathcal{S}^{c}_{y}(\ve{q},\omega)]/2$. 
 The total structure factor (i.e., FM and AFM chains combined), $\mathcal{S}_{\mu}(\ve{q},\omega)$ can be obtained similarly by summing over all four orbitals in the above Fourier transform. 

At zero magnetic field, the system exhibits magnetic order with ordering wavevectors $\ve{q} = [01] \equiv  (0,\pi)$ and $[21] \equiv (2\pi, \pi)$, corresponding to the FM and AFM chains ordered antiferromagnetically along the $\ve{\hat{y}}$ direction, respectively. 
This can be seen from the presence of the Bragg peaks in the static structure factor at low temperatures, as shown in Fig.~\subref{fig.dyn-sfac-temp}{(a)}. 
Total dynamical structure factor $\mathcal{S}(\ve{q},\omega)$ at a low temperature is plotted in Fig.~\subref{fig.dyn-sfac-temp}{(b)}. 
This spectrum compares favorably  with neutron scattering experiments~\cite{HZhang20}.
At our lowest temperature, $T = 3.074~\mathrm{K}$, we can resolve the $J_3 = 1.06~\rm{meV}\approx 12.3~\mathrm{K}$ interchain coupling. 
Above an energy scale set by $J_3$ we observe the two-spinon continuum of the AFM chains. 
Below this energy scale, the spinons bind to form spin-wave excitations. In particular, at the ordering wavevector $\ve{q} = [21]$ we observe large low-lying spectral weight (Bragg Peak as mentioned above), but are unable to resolve the spin-wave velocity. This scale is set by an effective coupling between the AFM chains that perturbatively tracks $(J_3)^2/J_2 = 0.212~\mathrm{meV}\approx 2.46$ K, and is hence smaller than the temperature scale of the simulation. Indeed, the linear spin-wave calculations show that the spin-wave velocity at [21] is very small (see the supplemental Material~\cite{SM}).
In the vicinity of $\ve{q} = [00]$, we observe a linearly dispersing spin-wave mode with a larger velocity, consistent with the linear spin-wave calculations~\cite{SM}. These gapless modes at [00] and [21] are associated with the spontaneous SU(2) spin-rotation symmetry breaking in the ordered ground state.

A standard route to dimensional reduction is through thermal fluctuations. As the temperature rises, thermal fluctuations first overcome the weak interchain coupling scale, suppressing interchain correlations, while intrachain correlations associated with the stronger intrachain couplings remain relevant. 
Consequently, the chains become effectively decoupled. 
Fig.~\subref{fig.dyn-sfac-temp}{(a)} shows the temperature evolution of the static structure factor. As the temperature increases, the peaks at $[01]$ and $[21]$ associated with magnetic ordering gradually weaken, and by $T\approx 12.3~\mathrm{K}\sim J_3$, their intensity is strongly suppressed. 
This signals the melting of the ordered state and the onset of effective chain decoupling. 
Figs.~\subref{fig.dyn-sfac-temp}{(b)–2(d)} show the corresponding dynamical structure factors. At $T\approx 12.3~\mathrm{K}$, the 
spectral weight associated with the linear spin-wave mode becomes strongly suppressed, while the two-spinon continuum extends to lower energies, consistent with the emergence of effectively decoupled AFM chains, albeit with substantial temperature  broadening.

 \begin{figure}
	\centering
        \includegraphics[width = \columnwidth]{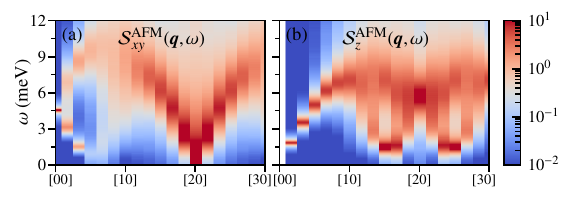}
        \caption[]{Dynamical structure factors of the decoupled AFM chains $($i.e., $J_3 = 0)$ at temperature $T = 3.074~ \mathrm{K}$ and magnetic field $B = 36.625~ \mathrm{T}$. Since $S_{\rm tot}^z$ is a conserved quantity, we leave [00] blank similar to Fig.~\ref{fig.dyn-sfac-temp}.} 
        \label{fig.decoup-chain}
\end{figure}

As mentioned above and as argued in Ref.~\cite{Reinold2025}, a different route to  dimensional reduction is through the application of an external magnetic field. Our  aim is to show that the dynamical structure factor of the model approaches that of a one-dimensional Heisenberg model in a magnetic field as the field strength increases. 
Before we perform simulations for the full model, we first consider the limit of decoupled chains by setting the interchain coupling $J_3=0$. 
At zero magnetic field, the dynamical structure factor of the decoupled AFM chains exhibits the well-known two-spinon continuum. The gapless mode of the continuum appears at $[20]$, rather than at $[10]$, due to the orbital positions of the AFM chain. 
At finite magnetic field, the spin chain develops a finite magnetization along the field direction, and the transverse and longitudinal dynamical structure factors become qualitatively different, as shown in Fig.~\ref{fig.decoup-chain}. 
The excitations remain gapless spinon excitations but  become incommensurate. In the transverse channel, an additional gapless mode appears near zero wavevector besides the one at $[20]$ [Fig.~\subref{fig.decoup-chain}{(a)}].  Furthermore, in the longitudinal channel, the magnetic field shifts the gapless mode from $[20]$ to a smaller wavevector [Fig.~\subref{fig.decoup-chain}{(b)}]. 
These shifts depend on the strength of the external field, with larger fields producing larger shifts. A complete analysis based on the Bethe Ansatz, explaining the field-induced reconstruction of the spinon continuum, can be found in Ref.~\cite{Muller1981}, and such field-dependent effects have been observed experimentally~\cite{Dender1997,Stone2003,Matsuda2017,Faure2019} and numerically~\cite{Grossjohann2009,Kohno2009}. 
Such field-dependent modifications are expected from the full model [Eq.~\eqref{eq.model}] if the proposed dimensional reduction holds. In the End Matter, we present a parton mean-field~\cite{Affleck88} description of the incommensurate spinon-excitations observed in the decoupled chain limit of Fig.~\ref{fig.decoup-chain}.

\begin{figure*}
        \centering
        \includegraphics[width = 0.95\textwidth]{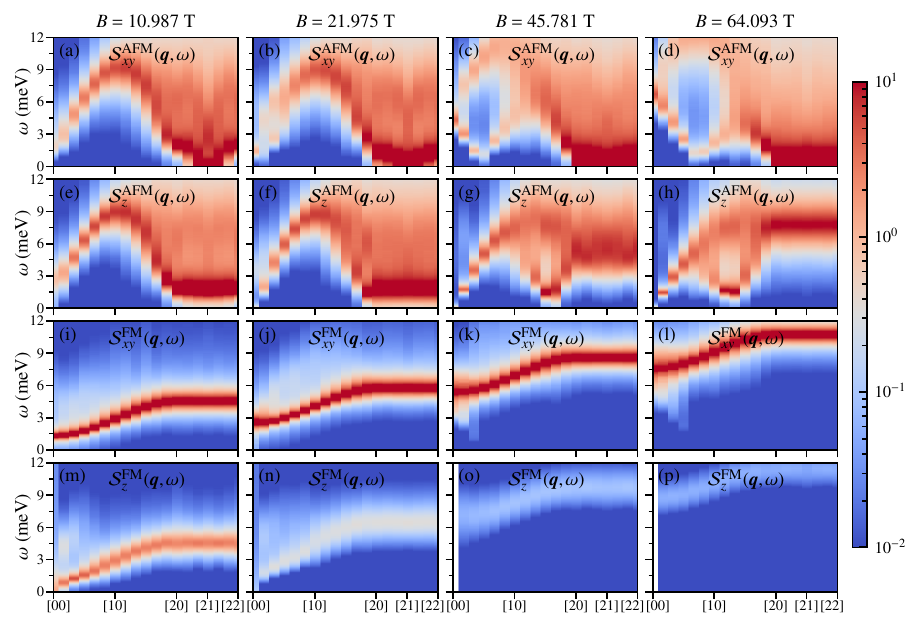}
        \caption[]{Separate contributions from the FM and AFM chains to the dynamical structure factor at temperature $T = 3.074 $K for various magnetic fields. 
        We leave [00] blank in (o) and (p) as the corresponding imaginary-time correlation functions are within the error bars.
        } 
        \label{fig.FM-AFM-sfac}
\end{figure*}

 \begin{figure}
	\centering
        \includegraphics[width = \columnwidth]{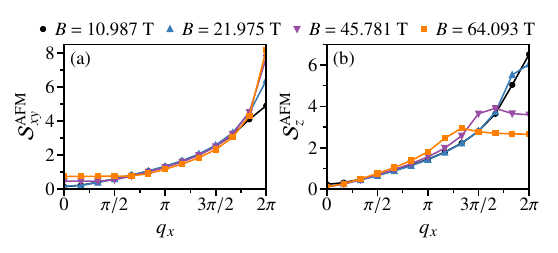}
        \caption[]{Static structure factor of the AFM chains for a path $(q_x,q_y) = (0,0) \rightarrow (2\pi,0)$ for several magnetic fields at $T = 3.074~ \mathrm{K}$. (a) Transverse and (b) longitudinal parts.} 
        \label{fig.AFM-stat-sfac}
\end{figure}

Next, we present the dynamical structure factors of the full model [Eq.~\eqref{eq.model}] for increasing magnetic fields in Fig.~\ref{fig.FM-AFM-sfac}. The separate contributions from the AFM chains are shown in Figs.~\subref{fig.FM-AFM-sfac}{(a)}--\subref{fig.FM-AFM-sfac}{(h)}, while those from the FM chains are shown in Figs.~\subref{fig.FM-AFM-sfac}{(i)}--\subref{fig.FM-AFM-sfac}{(p)}. 
The critical field beyond which the FM chains are fully polarized is $B_c \approx 16\,\rm{T}$~\cite{Reinold2025}. Below this field, the FM chains are not polarized yet, as reflected in the finite longitudinal spectral weight in Fig.~\subref{fig.FM-AFM-sfac}{(m)}. Additionally, the spinon continuum in the AFM transverse channel in Fig.~\subref{fig.FM-AFM-sfac}{(a)} remains gapped at $[20]$, indicating the persistence of interchain coupling in this regime. 
Beyond the critical field, the low-energy dynamics of the system approaches that of an AFM chain.  
This is confirmed by the numerical data at $B \approx 22\,\rm{T}$, where the two-spinon continuum extends to much lower energies (almost gapless) both in the transverse
[Fig.~\subref{fig.FM-AFM-sfac}{(b)}] and longitudinal [Fig.~\subref{fig.FM-AFM-sfac}{(f)}] channels. 
Furthermore, the dispersion becomes essentially flat along the $\ve{\hat{y}}$ direction [Fig.~\subref{fig.FM-AFM-sfac}{(f)}], indicating a strong suppression of propagation of AFM fluctuations across chains.  
In the FM chain contribution, the longitudinal spectral weight is strongly suppressed [Fig.~\subref{fig.FM-AFM-sfac}{(n)}], consistent with full polarization along the field. In the transverse channel, the spectral weight is shifted to higher energies [Fig.~\subref{fig.FM-AFM-sfac}{(j)}], reflecting the field-induced gap.  
This gap corresponds to the energy cost of flipping a spin, and the dispersion relation governs the propagation of the flipped spin (i.e., magnon). 
The stronger the field, the larger the gap, and the stiffer the FM chains against fluctuations. 
Consequently, at low energies, the FM chains become nearly ineffective at mediating correlations between neighboring AFM chains, as reflected in the pronounced flattening of the AFM dispersion perpendicular to the chains and the two-spinon continuum extending to very low energies.

While at $B\approx 22\,\rm{T}$ we observe clear dynamical signatures of dimensional reduction, the incommensurate field-induced character of the two-spinon continuum is not yet clearly apparent.   
By further increasing the field, for example to  $B \approx 45.8\,\rm{T}$ and $64.1\,\rm{T}$, we observe in Figs.~\subref{fig.FM-AFM-sfac}{(c)},  \subref{fig.FM-AFM-sfac}{(d)}, \subref{fig.FM-AFM-sfac}{(g)}, and \subref{fig.FM-AFM-sfac}{(h)} the incommensurate  field-induced features of an AFM spin chain, as previously shown in Fig.~\ref{fig.decoup-chain}. In the transverse channel, the spinon continuum splits and an additional gapless mode emerges near zero wavevector, whose position shifts to larger wavevectors with increasing field. 
Similarly, in the longitudinal channel the gapless mode moves from $[20]$ to a smaller wavevector, and this shift increases as the field is increased from $B \approx 45.8\,\rm{T}$ to $64.1\,\rm{T}$, in agreement with the behavior expected for a partially field-polarized AFM chain [Fig.~\ref{fig.decoup-chain}]. 
At the same time, the FM chain contribution in the transverse channel is pushed even higher in energy, and the longitudinal channel becomes essentially featureless [Fig.~\subref{fig.FM-AFM-sfac}{(p)}]. Thus, the FM chains are not able to fluctuate at low energies. Only the AFM chains remain dynamically active, but their fluctuations are effectively confined within each chain, as they cannot propagate from chain to chain due to the highly stiff FM chains in the middle. 

To further support the above dynamical signatures, we present the AFM chain contribution to the total static structure factor for various magnetic fields in Fig.~\ref{fig.AFM-stat-sfac}. 
As the magnetic field increases, the transverse part of the static structure factor increases around $[20]$, as shown in Fig.~\subref{fig.AFM-stat-sfac}{(a)}. This arises from the increasing spectral weight around $[20]$, associated with the two-spinon continuum [see Figs.~\subref{fig.FM-AFM-sfac}{(b)--4(d)}]. 
At smaller wavevectors, the transverse part also increases with field, reflecting additional spectral weight from the new field-induced low-energy mode near zero wavevector [see Figs.~\subref{fig.FM-AFM-sfac}{(b)--4(d)}]. 
In the longitudinal channel [Fig.~\subref{fig.AFM-stat-sfac}{(b)}], the structure factor peak progressively shifts away from the commensurate wavevector $[20]$ towards incommensurate wavevectors with increasing magnetic field. This behavior mirrors the evolution of the low-energy spectral weight from $[20]$ to incommensurate wavevectors as the field is increased [see Figs.~\subref{fig.FM-AFM-sfac}{(f)--4(h)}], and is characteristic of a partially field-polarized spin-$\frac{1}{2}$ AFM chain.

\textit{Discussion and Conclusions}---
We have shown that an SU(2) spin-symmetric unfrustrated model that is directly relevant for the mineral botallackite leads to distinct dynamical signatures of field-induced dimensional reduction. 
The key ingredient is the presence of alternating FM and AFM chains, 
and a hierarchy of scales $J_2 >J_1> J_3$ that ensures that in a magnetic field the FM chains polarize first. In this polarized phase, the transverse spin fluctuations of the AFM chains become effectively one-dimensional, as they cannot propagate from chain to chain. 
Our results show that this two-dimensional quantum magnet effectively maps onto an AFM Heisenberg spin-$\frac{1}{2}$ chain in a magnetic field, with gapless incommensurate spinon excitations at magnetic-field-dependent wavevectors. Hence, alongside the energy or temperature axis, a magnetic field can also drive a crossover from two- to one-dimensional behavior. This dimensional crossover can be directly probed through neutron scattering experiments on botallackite. Nevertheless, our modeling of botallackite is slightly simplified, since, e.g., it omits an exchange anisotropy that is dominant in the FM chain, as well as Dzyaloshinskii-Moriya interactions~\cite{HZhang20}. Although these interactions are smaller than the exchange couplings considered here, their inclusion could also lead to additional interesting physics. Provided that the orientation of the magnetic field is orthogonal to the exchange anisotropy, the ferromagnetic chain will be described by a transverse-field Ising model, and the transition to the polarized ferromagnetic chain falls into the two-dimensional Ising universality class. The $J_3$ coupling to the antiferromagnetic chains is such that the ferromagnetic chain perceives a longitudinal field. This opens the possibility of observing an $E_8$ particle spectrum \cite{Zamolodchikov89,Coldea10,Zhang2020,Amelin2020,Amelin2022}. In fact, recent experiments on botallackite~\cite{Reinold26} have been interpreted along these lines, and call for further numerical studies. 

\textit{Acknowledgments}---   
We would like to thank T. Sato and Z. Wang for useful discussions.  
We  gratefully acknowledge the Gauss Centre for Supercomputing e.V. for funding this project by providing computing time on the GCS Supercomputer SUPERMUC-NG at Leibniz Supercomputing,   (project number pn73xu) as  well  as  the scientific support and HPC resources provided by the Erlangen National High Performance Computing Center (NHR@FAU) of the Friedrich-Alexander-Universit\"at Erlangen-N\"urnberg (FAU) under the NHR project b133ae. NHR funding is provided by federal and Bavarian state authorities. NHR@FAU hardware is partially funded by the German Research Foundation (DFG) -- 440719683.
  F.F.A.\ acknowledges financial support from the DFG under the grant AS 120/16-1 (Project number 493886309) that is part of the collaborative research project SFB Q-M\&S funded by the Austrian Science Fund (FWF) F 86. 
 We also acknowledge financial support by the Deutsche Forschungsgemeinschaft (DFG, German Research Foundation) through the W\"urzburg-Dresden Cluster of Excellence {\it ctd.qmat} -- Complexity, Topology and Dynamics in Quantum Matter (EXC 2147, Project No.~390858490). S.K. also acknowledges financial support from the {\it ctd.qmat} FlexFund grant ``IN 0082025 SK''. J.v.d.B. acknowledges support from the Deutsche Forschungsgemeinschaft (DFG, German Research Foundation) within the Collaborative Research Center Correlated Magnetism: From Frustration to Topology (SFB 1143, Project No. 247310070).

\bibliography{./fassaad.bib}

\pagebreak
\onecolumngrid
\section*{End Matter}

\subsection{Parton Mean-Field Theory of the Incommensurate Spinon Excitations in the Heisenberg Antiferromagnetic Spin-$1/2$ Chain}
\begin{figure*}
    \includegraphics[width=\columnwidth]{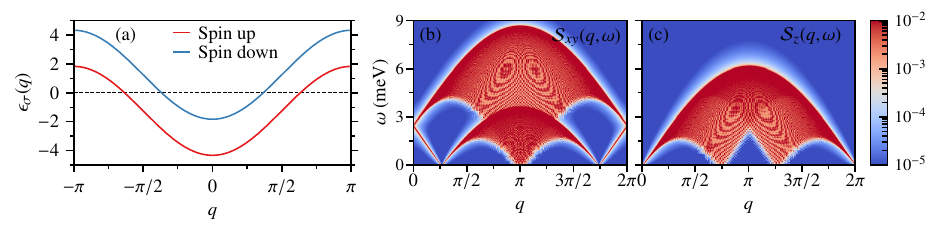}
    \caption{Antiferromagnetic Heisenberg spin-$1/2$ chain in the presence of a magnetic field. (a) Spinon dispersion for the self-consistent mean-field parameter $\chi = 1.541~{\rm meV}$ and magnetic field $B = 20~{\rm T}$.  (b) Transverse  and (c) longitudinal dynamical structure factors obtained from the parton mean-field analysis.} 
    \label{fig.parton}
\end{figure*}

We adopt the Abrikosov fermion representation of the spin-$\frac{1}{2}$ operators at $j^{\rm th}$ site of the chain, $\ve{S}_j \equiv \frac{1}{2} \ve{f}^{\dagger}_j
\ve{\sigma} \ve{f}_j$ with $\ve{f}^\dagger_j \equiv \left(f^\dagger_{j,\uparrow}, f^\dagger_{j,\downarrow}\right)$ a two-component fermionic spinor. To match the dimensions of the fermionic Hilbert space with the spin Hilbert space, we impose the constraint of single fermion occupancy per site, $\ve{f}_j^\dagger \ve{f}_j = 1$. In the fermionic Hilbert space, the antiferromagnetic Heisenberg chain Hamiltonian, $H = J \sum_j  \ve{S}_j \cdot \ve{S}_{j+1} - g\mu_{\rm B}B \sum_j S^z_j $, then takes the form: 
\begin{equation}
	 H= -\frac{J}{4} \sum_{b}\left( D_b^{\dagger} D_b + D_b D_b^{\dagger} \right)  -g\mu_{\rm B}B \sum_j 
   S^z_j + \frac{U}{2} \sum_i \left( \ve{f}_j^{\dagger} \ve{f}_j - 1 \right)^2.
\end{equation}
Importantly, the mapping is exact in the limit $U \rightarrow \infty$, where the constraint of single occupancy is imposed energetically, and $D_b \equiv \ve{f}_j^{\dagger} \ve{f}_{j+1}$ defined on the bond $b\equiv(j,j+1)$. To proceed,   we  use the Hubbard-Stratonovich transformation to decouple the four  fermion terms, and to obtain the action: 
\begin{equation}
S = \int_0^\beta d\tau \left[ \sum_j \ve{f}_j^{\dagger}(\tau)( \partial_\tau -i a_{0,j}) \ve{f}_j(\tau) + 
\sum_{b}  \frac{2|\chi_{b}|^2}{J} - \chi_b D_b^{\dagger} - \bar{\chi}_b D_b -g\mu_{\rm B}B \sum_j 
   S^z_j \right]. 
\end{equation}
In the above, we have taken the limit $U \rightarrow \infty$.  The  parton mean-field theory is obtained by taking the saddle point of the action. We impose the constraint of single occupancy on average, and assume a mean-field Ansatz of $\chi_b = \chi$ for all bonds. The  resulting mean-field Hamiltonian is then given by: 
\begin{equation}
   H_{\rm MF} =  -\sum_{b} \left( \chi D_b^{\dagger} + \bar{\chi} D_b \right) - g\mu_{\rm B}B \sum_j 
   S^z_j
   \label{eq.mean-field-Ham}
\end{equation}
with the self-consistency condition $\chi = \frac{J}{2} \langle D_b \rangle$, obtained from the saddle point approximation. We can choose $ \chi$ to be real, since its phase can be absorbed into the fermion operators. The mean-field Hamiltonian [Eq.~\eqref{eq.mean-field-Ham}] is quadratic in the fermions, and can be solved exactly. However, the mean-field parameter $\chi$ needs to be found self-consistently. The spinon bands are given by
\begin{align}
\epsilon_{\uparrow}(q) = -2\chi\cos(q)-\frac{g\mu_{\rm B}B}{2}, \,\,\,\epsilon_{\downarrow}(q) = -2\chi\cos(q)+\frac{g\mu_{\rm B}B}{2}.
\label{eq.spinon-bands}
\end{align}
For $J = 5.3~{\rm meV}$, $B = 20~{\rm T}$, and $\beta = 20~{\rm meV}^{-1}$, we obtain the self-consistent value $\chi = 1.541~{\rm meV}$. The spinon bands with this mean-field value are shown in Fig.~\subref{fig.parton}{(a)}, consisting of field-shifted  spin up and spin down bands.
The  spin susceptibility $\chi_{\mu}(q,\omega )   =  i \int_{0}^{\infty }  dt e^{i\omega t} \langle [S^{\mu}(q,t), S^{\mu}(-q,0)] \rangle$  with $S^{\mu}(q) = \frac{1}{\sqrt{L}} \sum_{j} e^{i q j} S^{\mu}_{j}$  corresponds to the particle-hole excitations of spinons, and  reads: 
\begin{equation}
	\chi_{z}(q,\omega) = \frac{1}{4L} \sum_{p,\sigma} \frac{f(\epsilon_{\sigma}(p + q)) - f(\epsilon_{\sigma}(p))}{ \epsilon_{\sigma}(p)  - \epsilon_{\sigma}(p + q) - \omega - i0^{+} },
\end{equation}
and 
\begin{equation}
	\chi_{xy}(q,\omega)\equiv \frac{\chi_x(q,\omega) + \chi_y(q,\omega)}{2}= \frac{1}{4L} \sum_{p,\sigma} \frac{f(\epsilon_{-\sigma}(p + q)) - f(\epsilon_{\sigma}(p))}{ \epsilon_{\sigma}(p)  - \epsilon_{-\sigma}(p + q) - \omega - i0^{+} }.
\end{equation}
Here $f(\epsilon)$ is the Fermi-Dirac distribution function, and $\epsilon_{\sigma}(p)$ is the spinon dispersion for spin $\sigma$, as given in Eq.~\eqref{eq.spinon-bands}. The transverse and the longitudinal dynamical structure factors are found from the above spin susceptibilities as $\mathcal{S}_{xy}(q,\omega) = \frac{1}{\pi}\frac{\mathrm{Im} \chi_{xy}(q,\omega)}{1-e^{-\beta \omega}}$ and $\mathcal{S}_{z}(q,\omega) = \frac{1}{\pi}\frac{\mathrm{Im} \chi_{z}(q,\omega)}{1-e^{-\beta \omega}}$. These dynamical structure factors at an infinitesimally small temperature are shown in Figs.~\subref{fig.parton}{(b)} and \subref{fig.parton}{(c)}.
Since $\chi_{z}$ ($\chi_{xy}$) accounts for scattering between the same (opposite) spin bands, distinct gapless modes appear at different wavevectors in the two channels, consistent with the QMC results of Fig.~\ref{fig.decoup-chain}.  In particular, these wavevectors are determined by the Fermi wavevectors of the spinon bands, which are in turn determined by the magnetization of the chain. While  this approximation captures the qualitative features of the incommensurate spinon excitations, it does not capture the full spectral weight distribution of the two-spinon continuum~\cite{Raczkowski2013}.

\clearpage

\addtolength{\oddsidemargin}{-0.75in}
\addtolength{\evensidemargin}{-0.75in}
\addtolength{\topmargin}{-0.725in}

\newcommand{\addpage}[1] {
\begin{figure*}
  \includegraphics[width=8.5in,page=#1]{supp_mat.pdf}
\end{figure*}
}

\addpage{1}
\addpage{2}
\addpage{3}
\addpage{4}
\addpage{5}
\addpage{6}

\end{document}